\documentclass[aps,pra,reprint,superscriptaddress,showkeys,nofootinbib,floatfix
]{revtex4-1}

\usepackage{articlelayout}
\usepackage{physmaths}
\usepackage{physunits}

\newcommand{\etal}{\textit{et al.}}
\newcommand{\figscale}{1}
\newcommand{\CF}{C$_4$F$_8$}
\newcommand{\CFx}[1]{CF$_{#1}$}
\newcommand{\SF}{SF$_6$}
\newcommand{\aSi}{\mbox{a-Si}}
\newcommand{\SiO}{SiO$_2$}
\newcommand{\rsf}{r_\text{SF6}}

\newcommand{\overetch}{f_\text{overetch}}

\newcommand{\SuppRefFigratevscoil}{\mbox{\ref{fig:ratevscoil}}} % fig:ratevscoil
\newcommand{\SuppRefFigprofiletrenchvssf}{\mbox{\ref{fig:profiletrenchvssf6}}} % fig:profiletrenchvssf6
\newcommand{\SuppRefTabcomp}{\mbox{\ref{tab:comp}}} % tab:comp
\newcommand{\SuppRefTabpassratevsplaten}{\mbox{\ref{tab:passratevsplaten}}} % tab:passratevsplaten
\newcommand{\SuppRefSecetchratesmeas}{\mbox{\ref{sec:etchratesmeas}}} % sec:etchratesmeas
\newcommand{\SuppRefSeceffectcoilpower}{\mbox{\ref{sec:effectcoilpower}}} % sec:effectcoilpower

\begin{document}
%%%%%%%%%%%%%%%%%%%%%%%%%%%%%%%%%%%%
% FRONTMATTER
\title{Inductively Coupled Plasma etching of amorphous silicon nanostructures over nanotopography using \texorpdfstring{C$_4$F$_8$/SF$_6$}{C4F8/SF6} chemistry}

% AUTHOR
\author{\surname{Harvey-Collard}, Patrick}
\email[Corresponding author: ]{P.Collard@USherbrooke.ca; {Tel.: +1 819-821-8000 x66204}; 
{Fax: +1 819-821-8046}}
\affiliation{Département de physique, Université de Sherbrooke, Sherbrooke, QC, J1K 2R1, Canada}
% AUTHOR
\author{\surname{Jaouad}, Abdelatif}
%\email[University email: ]{Abdelatif.Jaouad@USherbrooke.ca}
\affiliation{Département de génie électrique et de génie informatique, Université de Sherbrooke, Sherbrooke, QC, J1K 2R1, Canada}
% AUTHOR
\author{\surname{Drouin}, Dominique}
%\email[University email: ]{Dominique.Drouin@USherbrooke.ca}
\affiliation{Département de génie électrique et de génie informatique, Université de Sherbrooke, Sherbrooke, QC, J1K 2R1, Canada}
% AUTHOR
\author{\surname{Pioro-Ladrière}, Michel}
%\email[University email: ]{Michel.Pioro-Ladriere@USherbrooke.ca}
\affiliation{Département de physique, Université de Sherbrooke, Sherbrooke, QC, J1K 2R1, Canada}

% DATE
\date{September 20$^\text{th}$, 2012}

% ABSTRACT
\begin{abstract}
Inductively Coupled Plasma (ICP) etching of amorphous silicon (\aSi{}) nanostructures using a continuous \CF{}/\SF{} plasma over nanotopography in silicon dioxide (\SiO{}) is investigated. The coil power of the ICP system is used to tune the \aSi{} etch rate from $20$ to $125\nmpm$. The etch rates of \aSi{}, \SiO{} and electroresist are measured depending on the \SF{} ratio, platen power and chamber pressure and used to optimize the \aSi{}:\SiO{} etch selectivity. The results on nanostructures show that the presence of an insulating etch-stop layer affects the passivation ratio required to achieve vertical sidewalls. A low pressure is also necessary in order to etch the silicon nanostructure embedded into the oxide nanotrenches to form a highly conformable \aSi{} nanowire. We argue that both of these behaviors could be explained by surface charging effects. Finally, etching of $20\nm$ \aSi{} nanowires that cross $15\nm$ trenches in oxide with vertical sidewalls and a 4.3:1 \aSi{}:\SiO{} etch selectivity is demonstrated. This etching process can be used in applications where nanotopography is present such as single electron transistors or multigate transistors.
\end{abstract}

% KEYWORDS (6 max)
% Other possibilities: C4F8, SF6

% PACS possibles
% 52.77.Bn	Etching and cleaning (see also 81.65.Cf Surface cleaning, etching, patterning in surface treatments)
% 81.07.Gf	Nanowires
% 81.16.Nd	Micro- and nanolithography
% 85.35.Gv	Single electron devices
% 85.40.Hp	Lithography, masks and pattern transfer

\maketitle

%%%%%%%%%%%%%%%%%%%%%%%%%%%%%%%%%%%%
% ARTICLE

\section{Introduction}

Novel devices and applications in nanoelectronics require both high-resolution lithography and etching processes in order to pattern increasingly small features. The next-generation devices will use a variety of new materials and geometries, together with stringent scaling requirements. A large effort has been put in the development of plasma etching of nanostructures using a variety of gases and masks \cite{hung2011,welch2006,henry2010}. Nevertheless, there is still a need to address the specific problem of etching silicon nanostructures over nanotopography. Such capabilities are useful for the fabrication of Single Electron Transistors (SETs) with the nanodamascene process \cite{dubuc2008}, or poly-silicon multigate transistors \cite{ferain2011}. 

Plasma etching processes capable of patterning these nanostructures require a high etch selectivity of the etched material with respect to both the resist and the underlying etch-stop layer, a high anisotropy, the capability to etch into nanoscale topography and etch rates that allow good control of a very thin film etch. Achieving all these conditions simultaneously is challenging. It is also common that etching processes developed for larger structures fail when applied to nanometer-sized features, because they produce defects (i.e. scallops, roughness, trenching, footing \cite{franssila2010}) of comparable size to the actual features. These problems can be accentuated when nanostructures have to be etched on previously patterned ones.

This paper investigates the Inductively Coupled Plasma (ICP) etching of amorphous silicon (\aSi{}) nanostructures using \CF{}/\SF{} gases over previously patterned nanotopography in silicon dioxide (\SiO{}). The first series of experiments presented in Section \ref{sec:planarsubstrates} investigates the dependence of the etch rates and selectivities with the coil power, platen power, chamber pressure and gas concentration on unpatterned samples. Section \ref{sec:nanostructures} presents different etching parameters on samples with both nanotrenches and nanostructures, which points out the challenges of this type of etching. Then, we demonstrate the patterning of a small $20\nm$ \aSi{} line over and perpendicular to a nanotrench ($15\nm$ wide, $25\nm$ deep) in silicon dioxide with a $4.3$:1 silicon:oxide etch selectivity.

\section{ICP etching of planar substrates} \label{sec:planarsubstrates}

The etching mechanisms of the \CF{}/\SF{} plasma chemistry have been investigated for deep etching \cite{heinecke1975,dagostino1981,flamm1981,rueger1997,min2004,chen2007}. This section gives an investigation of the etch rates as a function of the \SF{} ratio, platen power and chamber pressure in a low power regime suitable for the patterning of shallow nanostructures. Then, the etching mechanisms of the mixed \CF{}/\SF{} plasma in this regime are discussed. This investigation allows the optimization of the etch selectivity of \aSi{} relative to oxide and \aSi{} relative to resist. Since the etch selectivity of \aSi{}:\SiO{} and \aSi{}:resist are found to be about the same and to follow the same tendencies, only the oxide one is emphasized because of its higher importance for the suggested applications. 

\subsection{Methodology}

Etch rates measurements are carried out by etching unpatterned samples of \aSi{}, \SiO{} and electroresist (\mbox{ma-N}). For the \SiO{} samples, the oxide is thermally grown ($1050\degC$) on a p-doped silicon substrate ($0.1-0.2\text{ }\Omega\,\text{m}$ resistivity). For the \aSi{} samples, the silicon is grown by Low Pressure Chemical Vapor Deposition (LPCVD) at $525\degC$ and $300\mTorr$ on the \SiO{} samples previously described. They are deoxidized in diluted hydrofluoric acid (HF $49 \%$:DI~water 1:50) immediately before etching. Throughout the article the term silicon is sometimes used instead of amorphous silicon, although the etched silicon is always amorphous and undoped. The \mbox{ma-N} samples are \SiO{} samples spin-coated with negative electroresist \mbox{ma-N} 2401 from Micro Resist Technology diluted 1:1 with anisole. The resist is baked at $90\degC$ for $1\minute$ and then at $100\degC$ for $10\minute$. Thicknesses of the \aSi{}, \SiO{} and \mbox{ma-N} layers are respectively $470\nm$, $107\nm$ and $35\nm$ and are measured with a spectroscopic ellipsometer before and after every etch. See the online supplementary material Section \SuppRefSecetchratesmeas{} for additional details.

The samples are etched in a STS Multiplex Advanced Silicon Etch ICP system. They are fixed on a quartz substrate with STI Crystalbond 555HMP adhesive for proper sample thermalization during the etch.  The ICP parameters investigated are the coil power $P_c$, the platen power $P_p$, the chamber pressure $p$, the flow ratio of \SF{} defined as $\rsf={(\text{SF}_6\text{ flow})}/{(\text{total flow})}$ and etch time $t$. The coil power is fixed to $100\Watt$ unless otherwise specified, the total gas flow is always $75\sccm$ and the gas used is a mixture of \CF{} and \SF{}. The platen temperature is $20\degC$. The coil and platen RF generators frequency is $13.56\MHz$.

\subsection{Effect of the ICP parameters on the etch rates and selectivities}

ICP reactors are designed to produce high density plasmas and mainly used for deep etching and high etch rates. Thus, to control the etching of shallow nanostructures, the etch rates must rather be decreased so as to insure a good process reproducibility and avoid plasma instabilities. The reactor capability to separate the plasma generation power (coil) from the ion acceleration power (platen) allow the etch rate to be tuned via the coil power. However, it can lead to a change in the dissociation ratio of the gas species and in the etching behavior. 

The etch rate measurements performed with coil powers ranging from $50\Watt$ to $600\Watt $ show that the etch rate can be tuned on a wide range of values, from $20\nmpm$ to $125\nmpm$ for amorphous silicon. More details are given in the online supplementary material (Section \SuppRefSeceffectcoilpower{}, Figure \SuppRefFigratevscoil{} and Table \SuppRefTabcomp{}). The changes in the plasma (ion and radical densities) induced by this unusually low power (the usual range of this reactor is $600-1000\Watt$) can be compensated by adjusting the gas ratio. 

A coil power of $100\Watt$ is chosen as an appropriate value and is used through the rest of the article. The chosen power is much lower than what is reported in the literature for similar systems. In comparison, Welch \etal{} \cite{welch2006} report powers in the range $600-1200\Watt$\footnote{Data obtained through personal communication with C. C. Welch.} and Hung \etal{} \cite{hung2011} of $800\Watt$. According to Henry \cite{henry2010a}, increasing the power in this regime reduces the etch rate of silicon, while Hung \etal{} \cite{hung2011} report that the etch rate is relatively insensitive to the coil power. These two statements are in disagreement with our experimental results, which show that increasing the coil power increases the etch rate. This difference could be explained by two factors. First, our regime is a low power one well suited for nanometric film etch, when the other reported work is at much higher power regimes. Second, because our ICP reactor is different, the power values used cannot be compared directly.

\begin{figure}[tbp]
   \centering
   \includegraphics[scale=\figscale]{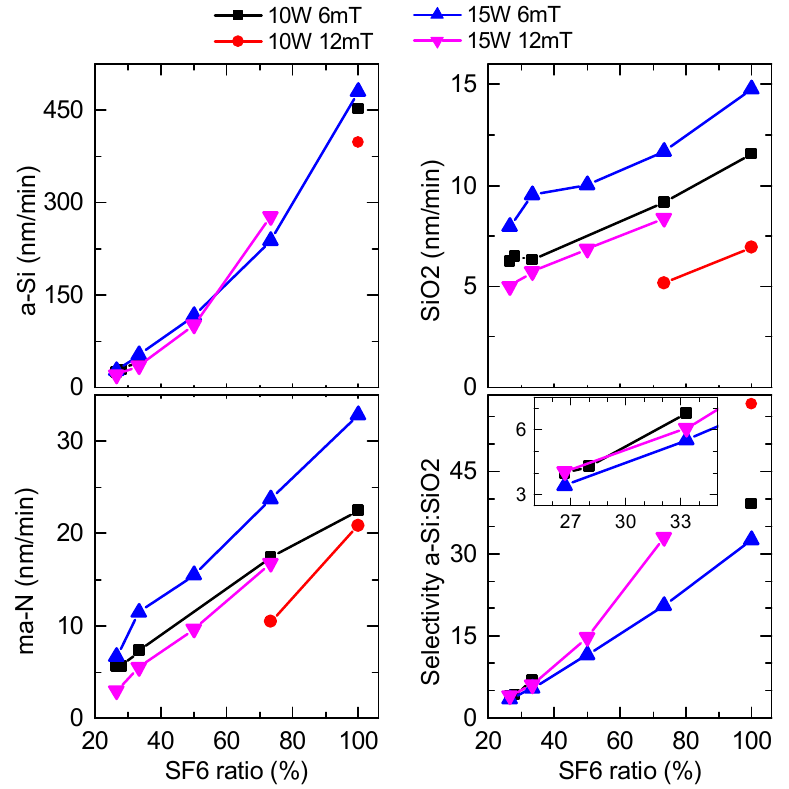} % requires the graphicx package
   \caption{Etch rates of silicon, oxide and resist depending on the \SF{} ratio. The silicon:oxide etch selectivity is calculated from the etch rates. The other ICP parameters are given in the legend in the order $P_p$ / $p$, where mT stands for mTorr. At very low \SF{} ratios ($\lesssim 20\pc$), the passivation dominates the etching and a polymer is deposited on the sample.}
   \label{fig:ratevssf6}
\end{figure}

The effect of the gas mixture is now examined. Figure \ref{fig:ratevssf6} 
% figure
% end 
shows the measured etch rates of silicon, oxide and electroresist as a function of the \SF{} ratio. The different curves in each plot are for different platen powers and chamber pressures. While the silicon etch rate variation is non-linear, the oxide and resist show linear dependencies. The etch selectivity of silicon with respect to its oxide is calculated from the etch rates and also shown in Figure \ref{fig:ratevssf6}. 

Silicon etching is caused by the neutral fluorine radicals released by \SF{} \cite{flamm1981} and this process is isotropic in our regime. Accordingly, the data of Figure \ref{fig:ratevssf6} shows that the highest silicon etch rates are obtained for pure \SF{}. When \CF{} is added to obtain anisotropy, the etch rate of silicon rapidly decreases. It is due to the presence of a steady-state passivation film that protects the silicon from being etched by fluorine, which has to diffuse through it to react with silicon \cite{rueger1997}. The steep slope indicates that too much passivation by \CF{} will highly degrade the \aSi{}:\SiO{} and \mbox{ma-N}:\SiO{} etch selectivity. It is important to note that \SF{} ratios of 20 to $40\pc$ are needed to obtain significant anisotropy. The optimal values reported in the literature \cite{welch2006,hung2011} depend on the ICP parameters. Our results using a variety of ratios are presented in the following section and in the online supplementary material Figure \SuppRefFigprofiletrenchvssf{}. 

For pure \CF{}, a polytetrafluoroethylene (PTFE)-like polymer is deposited on the surface. Our data in the online supplementary material Table \SuppRefTabpassratevsplaten{} shows that the deposition rate of this polymer is slightly enhanced by the platen power. This suggests that the passivation mechanism has an ionic component. 

It is clear from Figure \ref{fig:ratevssf6} that pure fluorine also etches the oxide. However, the etch rate reduction caused by the addition of \CF{} is much less dramatic than in the \aSi{} case. In addition, it is seen that the \SiO{} curve has a smaller slope than the \mbox{ma-N} one. This suggests that \CFx{x} components from \CF{} participate in the chemical etching reaction, an observation also reported in the literature \cite{flamm1981}. Carbon has a strong bond to oxygen and is abundant in the passivation film at the surface, whereas fluorine has to diffuse through the film. When carbon binds to oxygen to produce CO$_2$, more dangling bonds are left for fluorine to react with silicon. This process requires ionic bombardment to provide the activation energy \cite{lieberman2005}.

\begin{figure}[tbp]
   \centering
   \includegraphics[scale=\figscale]{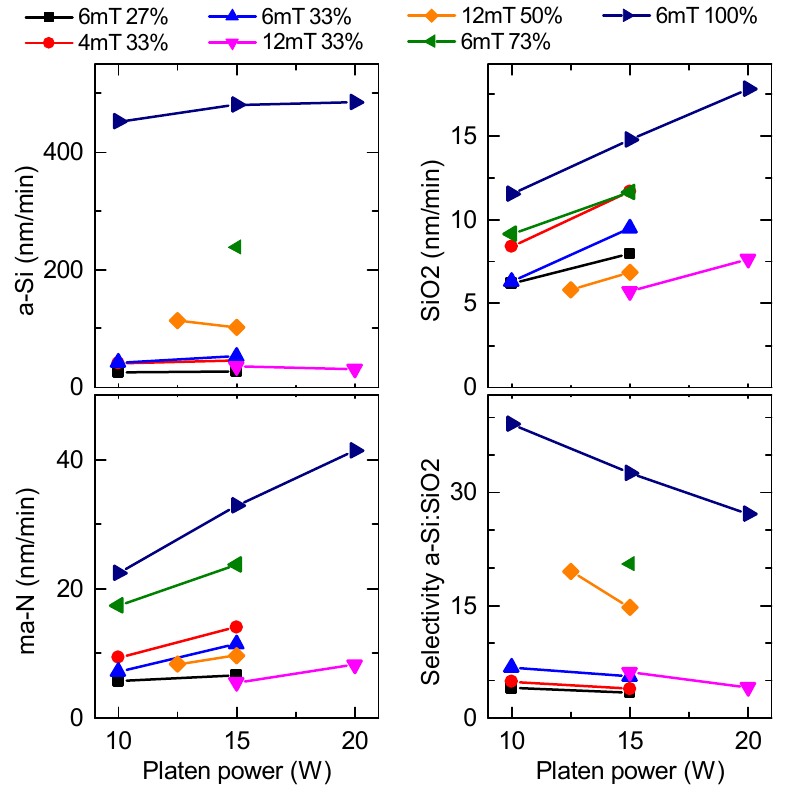} % requires the graphicx package
   \caption{Etch rates of silicon, oxide and resist depending on the platen power. The silicon:oxide etch selectivity is calculated from the etch rates. The other ICP parameters are given in the legend in the order \text{$p$ / $\rsf$}, where mT stands for mTorr.}
   \label{fig:ratevsplaten}
\end{figure}

The effect of the platen power is now examined. Figure \ref{fig:ratevsplaten} shows the measured etch rates as a function of the platen power. The latter causes the ions to be accelerated towards the sample. For silicon etching in pure \SF{}, the relatively small dependence of the silicon etch rate on platen power indicates that ions do not limit the etch rate (although they may play an activation role) \cite{min2004}. The Scanning Electron Microscope (SEM) inspection of etch profiles revealing isotropic etching (see online supplementary material Figure \SuppRefFigprofiletrenchvssf{} for images) reinforces this conclusion, because isotropy is typical of etching by neutral radicals. The relative insensitivity of the silicon etch rate to the platen power is not changed by the addition of \CF{}. However, anisotropy is observed due to the passivation of the sidewalls. The etch rates of \SiO{} and \mbox{ma-N} are dependent on the platen power, implying that their etching mechanisms are strongly ion-enhanced.
% figure
% end

\begin{figure}[tbp]
   \centering
   \includegraphics[scale=\figscale]{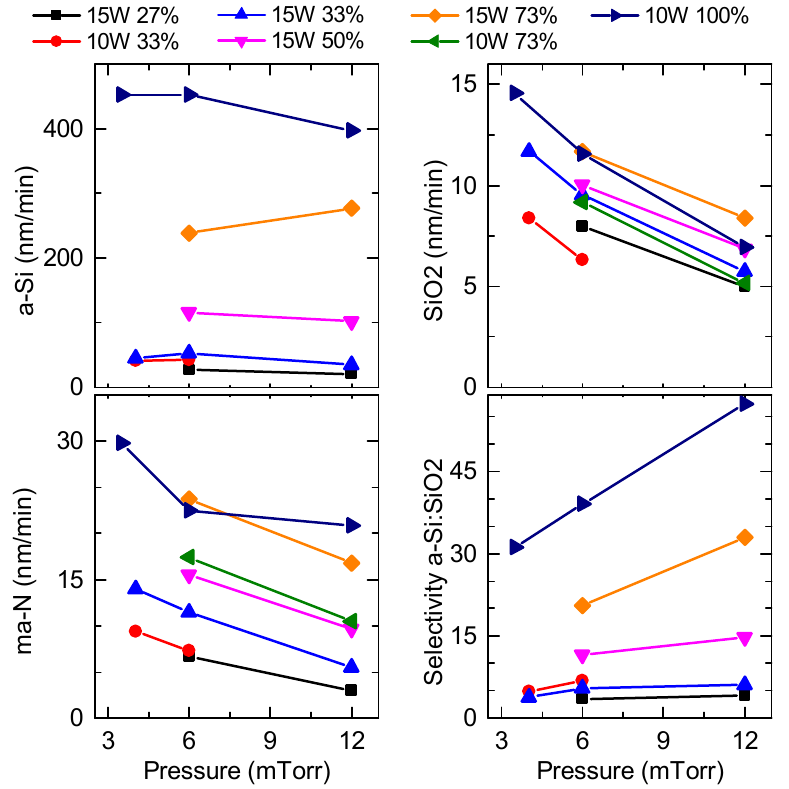} % requires the graphicx package
   \caption{Etch rates of silicon, oxide and resist depending on the chamber pressure. The silicon:oxide etch selectivity is calculated from the etch rates. The other ICP parameters are given in the legend in the order \text{$P_p$ / $\rsf$}.}
   \label{fig:ratevspressure}
\end{figure}

Figure \ref{fig:ratevspressure} plots the etch rates of silicon, oxide and resist as a function of the chamber pressure. Its effect is similar on the etch rates as the platen power one, namely that it affects mostly the oxide and the resist. Lowering the pressure is known to increase the flux and energy of the ions due to the longer mean free path and a modification of the sheath \cite{flamm1981}, but in different proportions than the platen power does. The silicon:oxide etch selectivity is also plotted in Figure \ref{fig:ratevspressure}.
% figure
% end

\section{ICP etching of nanostructures} \label{sec:nanostructures}

In this section, results related to nanostructures and nanotrenches are presented. The previous investigation of the dependence of the silicon:oxide and silicon:resist etch selectivity on the ICP parameters allows the process to be nearly optimal concerning selectivity, while exploring the impact of these parameters on other etching characteristics such as anisotropy, Aspect Ratio Dependent Etch rate (ARDE) or footing \cite{franssila2010}. We discuss the factors that influence the vertical profile of the etched structures and the capacity of the process to etch into nanotrenches. Finally, we present a working process and discuss the trade-offs of the different parameters.

\subsection{Methodology}

Three different types of samples are used to investigate the behavior of the process on both nanotrenches and nanostructures. Type A samples are patterned for a SET application (see Figures \ref{fig:ttobratio}b and \ref{fig:ttobratio}c). They have silicon nanowires patterned over trenches in oxide. Type B samples are thick \aSi{} samples patterned with negative \mbox{ma-N} resist test structures. Type C samples are thick \aSi{} samples patterned with positive ZEP resist. In the following paragraphs, we detail the fabrication parameters of the samples.

Type A samples have a $107\nm$ thermally grown \SiO{} layer. The oxide is plasma etched with a mask of positive ZEP electroresist to form $15-25\nm$ wide and $20\nm$ deep trenches in the oxide using the process developed by Guilmain \etal{} \cite{guilmain2011}. The trenches are then filled by a $40\nm$ thick \aSi{} LPCVD layer. A $35\nm$ thick \mbox{ma-N} mask is used for the \aSi{} nanowires that will be etched using the \CF{}/\SF{} plasma. The resist and polymer residue are ashed in an oxygen plasma after the etch (type A only).

Type B samples have a thick $470\nm$ \aSi{} layer over the $107\nm$ \SiO{} samples prepared as described in the previous section. A \mbox{ma-N} mask is used for the \aSi{} nanostructures.

Type C samples are similar to type B, but are patterned with positive ZEP resist instead. The Zeon Corp. ZEP 520A resist is diluted 2.4:1 (weight) with anisole and its thickness is $90\nm$. All three types of samples are deoxidized in diluted HF with the patterned resist prior to the etch.

For the type A samples, the etch time $t$ is calculated as following. Knowing the etch rates of planar films, the etch time of the native oxide, \aSi{} bulk film and trench depth is calculated and added. The total time is then multiplied by an overetch factor $\overetch$ chosen between $1.10$ and $1.25$. For the type B and C samples, the etch time is chosen to create either structures or trenches $30-100\nm$ deep.

\subsection{Etching in small nanotrenches}

\begin{figure}[tbp]
   \centering
   \includegraphics[scale=\figscale]{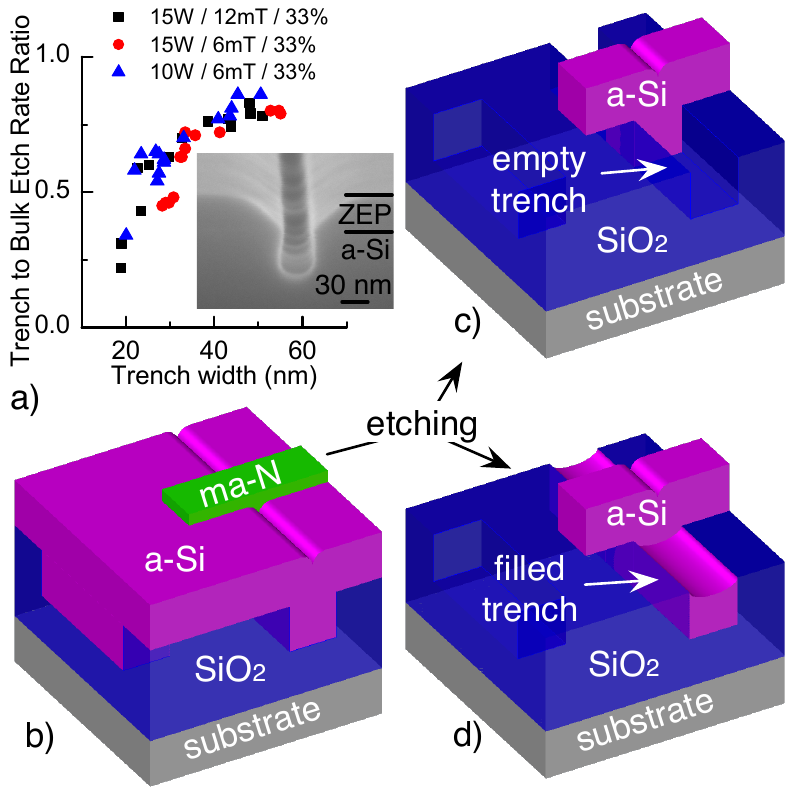} % requires the graphicx package
   \caption{a) Etch rate ratio between narrow trenches in \aSi{} and wide $10\um$ trenches (type C samples). The ICP parameters are given in the legend in the order \text{$P_p$ / $p$ / $\rsf$}. In this test, trenches are etched in \aSi{} using ZEP electroresist as a mask (SEM micrograph inset). All the tested parameters seem to have the same behavior when etching in narrow ZEP and \aSi{} trenches. b) Schematic of type A samples before etching. c) Schematic of type A samples after etching. This is the target structure. d) For some parameters, the ICP process is unable to etch the \aSi{} inside the nanotrench, as illustrated on the schematic.}
   \label{fig:ttobratio}
\end{figure}

A reduction of the etch rate in small trenches depending on their aspect ratio, a phenomenon known as ARDE, is expected due to microloading. In Figure \ref{fig:ttobratio}a, the etch rate of \aSi{} trenches defined by an opening in positive ZEP resist is measured and compared to the rate of a $10\um$ opening considered to be the bulk etch rate. Three processes are tested with different pressures and platen powers, but same \SF{} ratios, to check for the best parameters to reduce ARDE. The data shows that there is no significant difference between the three.
% figure
% end

\begin{figure}[tbp]
   \centering
   \includegraphics[scale=1]{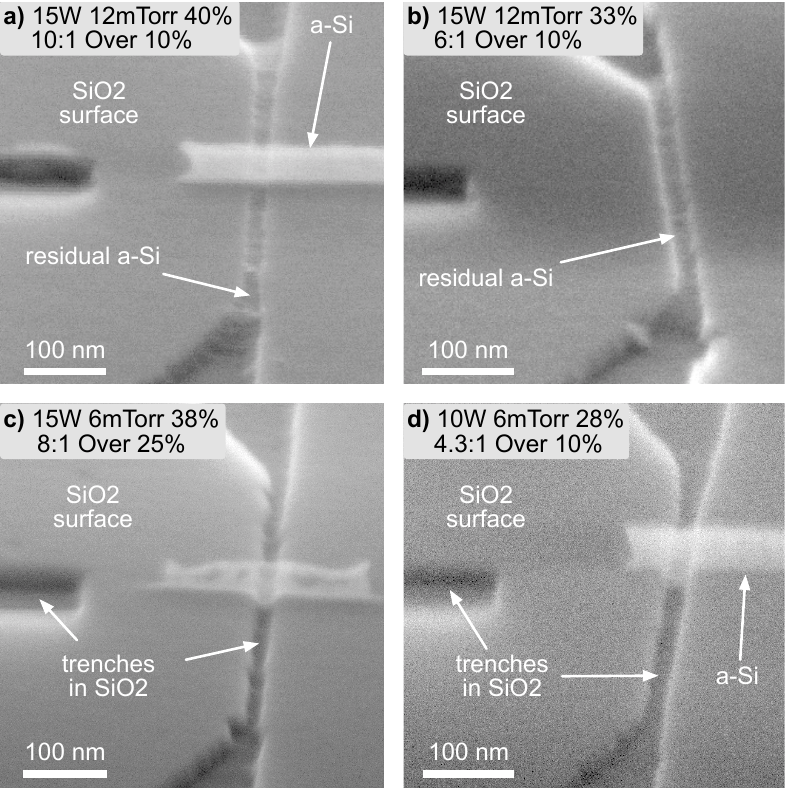} % requires the graphicx package
   \caption{Effect of different etch parameters on the etching of silicon inside small ($<25\nm$) trenches in oxide (type A samples). The parameters are given in the order $P_p$ / $p$ / $\rsf$ / \mbox{\aSi{}:\SiO{}} etch selectivity / overetch. a) and b) resemble the schematic of Figure \ref{fig:ttobratio}d. A good process is presented in d), where we see that the \aSi{} nanowire goes down inside the trench like in Figure \ref{fig:ttobratio}c.}
   \label{fig:trenchvspressure}
\end{figure}

Similar parameters are then tested on type A samples, which are the target structures illustrated in Figure \ref{fig:ttobratio}c. In Figure \ref{fig:trenchvspressure}, SEM images of the etched nanostructures are shown. We see that in Figures \ref{fig:trenchvspressure}a and \ref{fig:trenchvspressure}b the oxide trench is not empty, while the ones of Figures \ref{fig:trenchvspressure}c and \ref{fig:trenchvspressure}d are. Hence, the same three conditions used for the test of Figure \ref{fig:ttobratio}a are not equivalent in this situation. Other structures etched with $25\pc$ overetch show the same results. Comparing Figures \ref{fig:trenchvspressure}a with \ref{fig:trenchvspressure}c, and \ref{fig:trenchvspressure}b with \ref{fig:trenchvspressure}d, we find that the successful conditions are the $6\mTorr$ pressure ones, a result that could not be anticipated from the test of Figure \ref{fig:ttobratio}a. Other combinations of parameters were tested and confirm these results. In particular, a $10\Watt$ / $6\mTorr$ / $33\pc$ / 7:1 / Over$\,10\pc$ process having about the same etch selectivity and physical etching as Figure \ref{fig:trenchvspressure}b (because their oxide and resist etch rates are similar and are mostly of ionic origin) was tested, yet only the $6\mTorr$ process clears the \aSi{} in the trench. All the nanowires show undercutting because of the high \SF{} ratio, except for the one in Figure \ref{fig:trenchvspressure}d. This last process demonstrates a good vertical profile and a trench clear of silicon, which is the intended result.
% figure
% end

\subsection{Anisotropy and surface effects}

Anisotropy is obtained by passivating the sidewalls with a fluorocarbon polymer. In Section \ref{sec:planarsubstrates}, we showed that \CF{} provides this passivation at the expense of the silicon:oxide etch selectivity. Therefore, tuning the \SF{} ratio is important for both anisotropy and etch selectivity. 

\begin{figure}[tbp]
   \centering
   \includegraphics[scale=1]{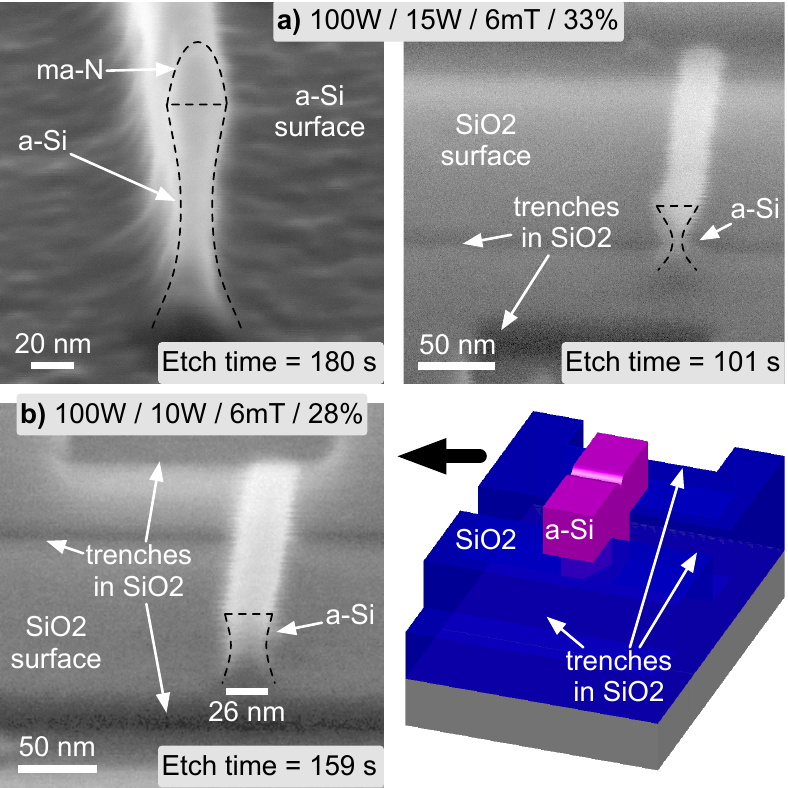} % requires the graphicx package
   \caption{SEM sideview of an \aSi{} nanowire for different etch-stop layers. a) In the left image, there is no stop layer (type B samples, surface is \aSi{}). In the right image, the etch is carried out over topography: the etch stops on the oxide layer (type A samples, surface is \SiO{}). The silicon nanowire is more severely undercut in the presence of the oxide stop layer. b) Good vertical profile is realized by using a lower \SF{} ratio.}
   \label{fig:profilevssurface}
\end{figure}

In Figure \ref{fig:profilevssurface}a, 
% figure
% end
type A (right) and type B (left) samples are etched and imaged with an SEM at a low $15\deg$ tilt angle. In the left image, there is no stop layer, so the surface exposed to the plasma is \aSi{}. In the right image, the etch is carried out over topography and the etch stops on the oxide layer (surface is \SiO{}). Because of the need to etch into the nanotrench, almost $50\pc$ of the etch is carried out while the oxide surface is exposed to the plasma. The processes are underpassivated, leading to some undercutting of the features. Nevertheless, the structure on the right is much more undercut than the one on the left, even if the etch time is smaller. Most of the right sample structures were in fact completely collapsed or about to. This phenomenon was observed with other sets of parameters as well. 

The authors in references \cite{welch2006,hung2011} report that \SF{} ratios of $30-35\pc$ produce vertical sidewalls and that the optimal value for verticality is pressure-dependent. Although in our work such values also produced near-vertical sidewalls when using type B samples, lower $26-28\pc$ values were needed for type A samples because of the oxide layer, as shown in Figure \ref{fig:profilevssurface}b. According to Figure \ref{fig:ratevssf6}, this reduces the \aSi{}:\SiO{} etch selectivity by about a third. Since it was previously established that the pressure is the key to clearing the silicon from the nanotrenches, the platen power can be lowered from 15 to $10\Watt$ with no impact on the silicon nanotrench etching. According to Figure \ref{fig:ratevsplaten}, a lower platen power allows to compensate partly for the drop in selectivity caused by the lower \SF{} ratio. Hence, our work shows that the optimal \SF{} ratio also depends on the presence or absence of an oxide etch-stop layer and provides a method for optimizing the processing conditions. See the online supplementary material Table \SuppRefTabcomp{} for a comparison of the etching parameters for the two different scenarios.

\subsection{Discussion on the surface charging effects}

We demonstrated that the presence of the oxide surface increases the undercut on standing \aSi{} nanostructures (Figure \ref{fig:profilevssurface}) and that it inhibits the etching into trenches when using a high pressure process (Figure \ref{fig:trenchvspressure}). A possible explanation for these results is that the oxide surface charges positively when exposed to the plasma. Charging effects can cause sidewalls to be undercut, provoke footing and can slow the etching of surfaces \cite{arnold1991,park2003}. In our case, the effect observed has to be related to the oxide surface, because any effect caused by the nanostructure would always be present and thus unnoticed. A positive charge on the oxide would deflect the ions towards the sidewalls, causing etching of the passivation layer and an undercut as observed. It could also repel ions with low energy, preventing them from clearing the passivation layer at the surface of the trench.

The results show that the pressure is the parameter that has the most impact on the nanotrench etching, and similar results are reported by Park \etal{} \cite{park2003}. We suggest that a lower pressure changes the energy distribution of the ions towards higher energies, while changing the amount of charge build-up on the surface, in a way that is favorable to overcome the charging effects. This would be consistent with the work of Park \etal{} \cite{park2003}. 

We also see that using a higher platen power does not enhance the etching into the oxide trenches, because processes with $P_p=15\Watt$ work only at low pressures. Moreover, high platen powers degrade the \aSi{}:\SiO{} etch selectivity. Also, in the online supplementary material (Table \SuppRefTabpassratevsplaten{}), we show using pure \CF{} that increasing the platen power increases the deposition rate of the fluorocarbon polymer. This suggests that the passivation layer is itself charged because of the ionic contribution. 

We also exclude that the problem observed is an effect of the \aSi{}:\SiO{} etch selectivity. In Figure \ref{fig:trenchvspressure}, we see no correlation between etch selectivity (4:1 to 10:1) and the capacity to etch inside the trenches. Nevertheless, we observed in other experiments that using an extremely high \SF{} ratio ($73\pc$) enables the trench etching at high pressures. In this ratio regime, the etch selectivity has a high 34:1 value but the etching is isotropic. We suggest that in this extreme case, the steady-state passivation layer on the surface is not sufficient to prevent the neutral fluorine from etching the trench, even in the case of charge build-up.

Experiments with a precise monitoring of the density and energy of the ions, combined with simulation, would be needed to confirm these predictions, but are beyond the scope of this work.

\section{Conclusion}

ICP etching of amorphous silicon nanostructures over nanotopography has been realized using \CF{}/\SF{} continuous plasma. Our results show that the coil power can be used to tune the etch rate to allow reproducible etching of $40\nm$ thin films. Using a power of $100\Watt$, we have then demonstrated how the etch rates and selectivities depend on the \SF{} ratio, platen power and chamber pressure. These results give insights on the microscopic etching mechanisms. Our processes have been applied to various types of nanopatterned samples. We have observed that the amount of passivation by \CF{} needed to obtain vertical sidewalls is higher in the presence of an oxide etch-stop layer. To etch the silicon inside the oxide nanotrenches, it is necessary to use a low pressure of $6\mTorr$. While a $12\mTorr$ pressure or higher allows a better \aSi{}:\SiO{} etch selectivity, the etch stops when the oxide surface is exposed, leaving the trench incompletely etched, even with a severe overetch. It has been argued that the surface-dependent vertical profile and the nanotrench etching can both be explained by charging effects. Finally, we have demonstrated a process to produce highly conformable $20\nm$ \aSi{} nanowires that cross $15\nm$ trenches in oxide with vertical sidewalls and an \aSi{}:\SiO{} etch selectivity of 4.3:1 suitable for applications with nanotopography, such as multigate transistors or SETs.

\begin{acknowledgments}
The authors thank C. Sarra-Bournet for fruitful discussions concerning this work and C. Bureau-Oxton for the careful reading the manuscript. This work was supported by NSERC, FRQNT, NanoQuébec and CIFAR. 
\end{acknowledgments}

\bibliography{References}
\bibliographystyle{apsrev-no-issn}

\clearpage

% !TEX root = Manuscript.tex
%%%%%%%%%%%%%%%%%%%%%%%%%%%
%%%%% SUPPLEMENTARY MATERIAL %%%%%

% Article references
% Figures
\newcommand{\RefFigratevssf}{\mbox{\ref{fig:ratevssf6}}} % fig:ratevssf6
\newcommand{\RefFigttobratio}{\mbox{\ref{fig:ttobratio}}} % fig:ttobratio

% Numéros de section pour Supplementary
%\setcounter{section}{0}     % Pour resetter les numeros de section
% Numéros de figure pour Suplementary
\makeatletter
\renewcommand\thefigure{\mbox{S-\arabic{figure}}}
\makeatother
%\setcounter{figure}{0}     % Pour resetter les numeros de figure
% Numéros de tableaux pour Supplementary
\makeatletter
\renewcommand\thetable{\mbox{S-\Roman{table}}}
\makeatother

\newpage
\onecolumngrid
{\centering
\large\textbf
{Online Supplementary Material for: Inductively Coupled Plasma etching of amorphous silicon nanostructures over nanotopography using C$_4$F$_8$/SF$_6$ chemistry} \\ \rule{0pt}{12pt}
} 
\twocolumngrid

\section{Details on etch rates measurements} \label{sec:etchratesmeas}

To perform the etch rate measurements, the thicknesses of the \aSi{}, \SiO{} and ma-N layers are measured with a spectroscopic ellipsometer before and after every etch. The etched thickness is then divided by the etch time to compute the etch rates. The etch selectivity is defined as the etch rate ratio of two materials.

The measured ellipsometric thicknesses are consistent with profilometric measurements. The \aSi{} layer includes a native oxide layer. The refractive index $n$ and absorption coefficient $k$ of the ellipsometric model of the \aSi{} and ma-N layers is fitted along with the thickness every time. The $n$ and $k$ did not change significantly after the etch. All other layers have a fixed $n$ and $k$ value. Some roughness and fluorocarbon etch residue is seen in the model as a higher native oxide thickness after the etch, which is typically $1-2.5\nm$. All fits are good and fitting error is low.

The measurements are accurate within one nanometer and the etched thicknesses range from 25 to $400\nm$ depending on the etch rate. This gives a relative uncertainty of $0.5\pc$ to $8\pc$ in the worst case. The variability in the etch rate measurements is also due to the etch rate increasing slightly over time. The etch times range from $45\s$ to $120\s$. For instance, the data point at $33\pc$ \SF{} of the $15\Watt$ / $6\mTorr$ data set of Figure \RefFigratevssf{} has an etch time of $120\s$, whereas the others have a $60\s$ one. This point is off with the others by $15\pc$ because the etch rate increased during the etch. Despite these measurement imprecisions, it’s the dependency of the etch rates with the ICP parameters that's important. Having extremely precise values is of secondary importance since they would be different on a different reactor. Error bars do not appear in the article because they make the graphs harder to read.

\section{Effect of the coil power} \label{sec:effectcoilpower}

In this section, more data is presented on the effect of the coil power on the etch rates. Since the main objective of the article is to investigate the etching of nanostructures, the effect of the coil power has not been studied  thoroughly. It is sufficient to find a regime where the total etch time allows the reproducible etching of a very thin  film (about $40\nm$). The consistency of the results presented in the article is a strong indication that the plasma is stable even in the low power regime used ($100\Watt$) and can be used to etch nanostructures reliably.

Using the coil power, the etch rate of \aSi{} can be varied from about 20 to $125\nmpm$. These values are taken from different etching processes used throughout the paper with a fixed $26-28\pc$ \SF{} ratio (but with no specific chamber pressure or platen power). This gas ratio allows a high anisotropy and almost vertical sidewalls. The conclusion of all the tests conducted is that the coil power has a high impact on the etch rates, while the other properties are only slightly affected. This impact has not been studied quantitatively.

\begin{figure}[tbp]
   \centering
   \includegraphics[scale=1]{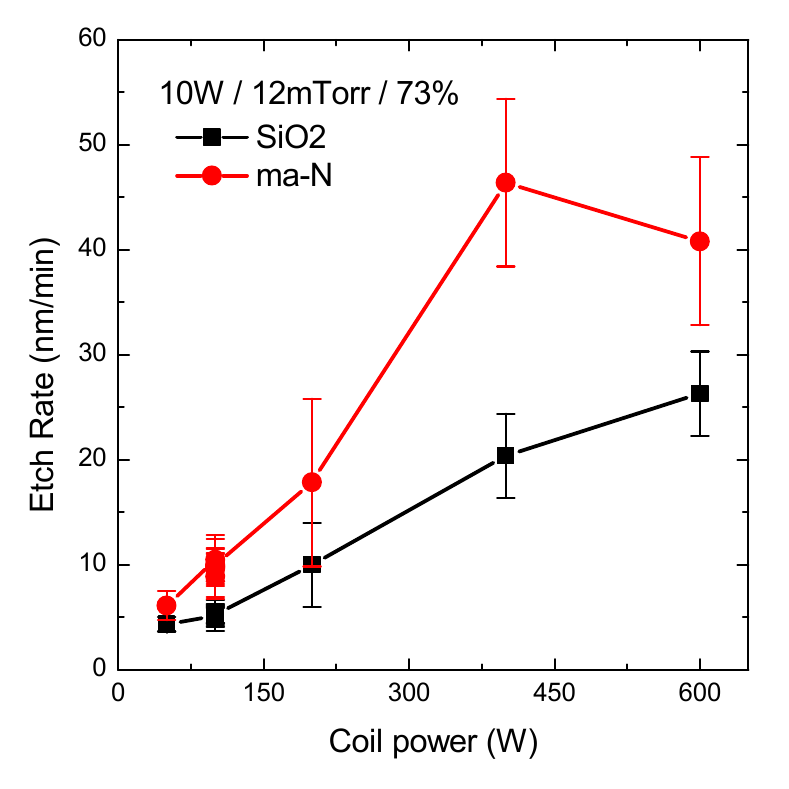} % requires the graphicx package
   \caption{Etch rate of silicon dioxide and ma-N resist depending on the coil power. The other ICP parameters are given in the legend in the order $P_p$ / $p$ / $\rsf$.}
   \label{fig:ratevscoil}
\end{figure}

As an exemple of such a coil power effect, Figure \ref{fig:ratevscoil} plots the oxide and resist etch rates as a function of the coil power. With $73\pc$ \SF{}, this process is nearly isotropic. The short $15-45\s$ etch times explain the high error bars on the data. The erratic point on the ma-N curve at $400\Watt$ is due to the etch time of $15\s$, which is too short to allow a good plasma stability. The \aSi{} data set is missing because the \aSi{} thin films used were too thin ($40\nm$, which is thinner than those of the data sets of the article) and the \aSi{} etch rates too fast, meaning that they were completely etched. 
% figure
% end

In Table \ref{tab:comp}, we give details about two similar processes at different coil powers. The first one is tuned for an application that doesn't require a lot of overetching after an oxide surface is reached. The second is tuned for an application where nanotopography is involved and significant overetching is required, like the one of Figure \RefFigttobratio{}. Because a lower pressure is necessary, the etch selectivity is reduced by almost a factor of 2.
% table
\begin{table}[tbp]
\caption{Comparison between achievable selectivities when etching bulk \aSi{} and over topography in oxide. The topography requirement cost almost a factor of 2 in etch selectivity because of the lower pressure required. Note that it is not the lower coil power used that causes this loss in etch selectivity.}
\begin{center}
\begin{tabular}{lll}
	\hline \hline
	Process & Normal &Over topography \\
	\hline
	$P_c$ (W) & 600 & 100 \\
	$P_p$ (W) & 10 & 10 \\
	$p$ (mTorr) & 12 & 6 \\
	\SF{} ratio (\%) & 26.7 & 28.0 \\
	\aSi{} (nm/min) & 125 & 27.9 \\
	\aSi{}:\SiO{} & 7.5:1 & 4.3:1 \\
	\hline
\end{tabular}
\end{center}
\label{tab:comp}
\end{table}
% end

\section{Passivation rate}

In Table \ref{tab:passratevsplaten}, the deposition rate of the fluorocarbon polymer deposited by \CF{} on a \SiO{} sample is measured. The passivation rate is enhanced by the platen power and hence has an ionic component.
% table
\begin{table}[tbp]
\caption{Fluorocarbon passivation layer deposition depending on platen power. The other ICP parameters are $P_c=100\Watt$, $p=12\mTorr$, $\text{C}_4\text{F}_8\text{ flow}=20\sccm$.}
\begin{center}
\begin{tabular}{ll}
	\hline \hline
	Platen Power (W) & Deposition Rate (nm/min) \\
	\hline
	20 & 26 \\
	50 & 32 \\
	\hline
\end{tabular}
\end{center}
\label{tab:passratevsplaten}
\end{table}
% end

\section{Effect of the \SF{} ratio on the vertical profiles}

\begin{figure*}[tbp]
   \centering
   \includegraphics[scale=1]{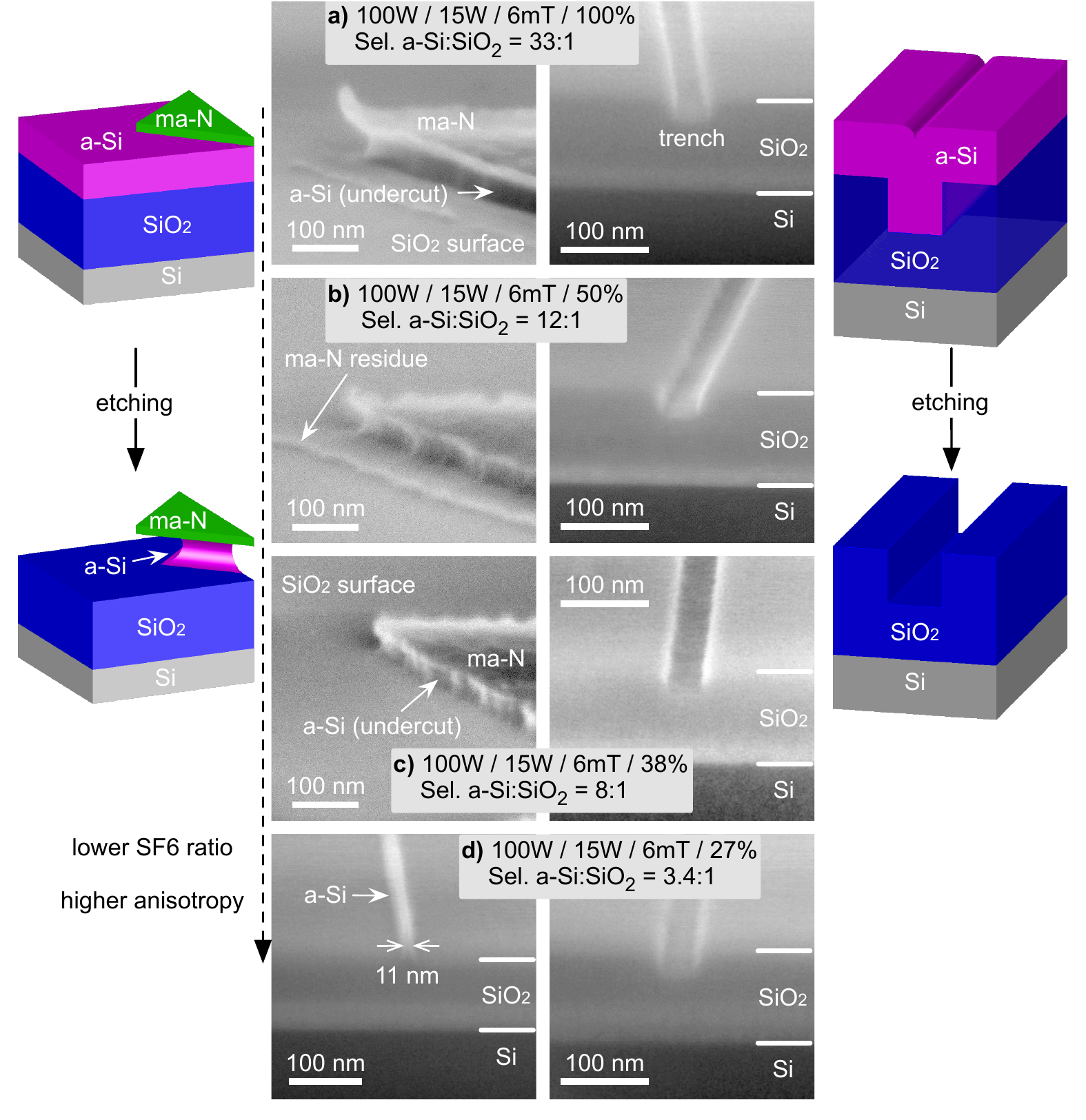} % requires the graphicx package
   \caption{Etch profile of type A samples over a wide range of \SF{} ratios. Left column: $40\nm$ thick \aSi{} structures. Right column: trenches in oxide. The parameters are given in the order $P_c$ / $P_p$ / $p$ / $\rsf$, where mT stands for mTorr. Trenches are about $40\nm$ wide and $26\nm$ deep. In a), b) and c), the \aSi{} structures are clearly undercut, showing too low or no passivation. In d), an $11\nm$ wide nanowire shows a vertical profile at the expense of the \aSi{}:\SiO{} etch selectivity. In all images the trenches in oxide are clear of silicon, an important property for etching over topography.}
   \label{fig:profiletrenchvssf6}
\end{figure*}

In Figure \ref{fig:profiletrenchvssf6}, the left column provides data showing the undercut of nanostructures depending on the \SF{} ratio. The etching is isotropic at $100\pc$ \SF{} and perfectly anisotropic at $27\pc$ \SF{}. This last setting allows the fabrication of small $11\nm$ nanowires. In the right column, trenches in silicon dioxide are cleared of their silicon filling, a characteristic that is required when patterning structures over topography.
% figure
% end

\clearpage

\end{document}